# fastMRI+: Clinical Pathology Annotations for Knee and Brain Fully Sampled Multi-Coil MRI Data

Ruiyang Zhao[1,2,3], Burhaneddin Yaman[1,4], Yuxin Zhang[1,2,3], Russell Stewart[1,5], Austin Dixon[1,6], Florian Knoll[7], Zhengnan Huang[7], Yvonne W. Lui[7], Michael S. Hansen[1], Matthew P. Lungren[1,5]

[1] Microsoft Research
[2] University of Wisconsin-Madison, Department of Radiology
[3] University of Wisconsin-Madison, Department of Medical Physics
[4] University of Minnesota, Department of Electrical and Computer Engineering
[5] Stanford University, School of Medicine
[6] Duke University, School of Medicine
[7] New York University, School of Medicine

Contact: Michael.Hansen@microsoft.com

**Abstract**

Improving speed and image quality of Magnetic Resonance Imaging (MRI) via novel reconstruction approaches remains one of the highest impact applications for deep learning in medical imaging. The fastMRI dataset, unique in that it contains large volumes of raw MRI data, has enabled significant advances in accelerating MRI using deep learning-based reconstruction methods. While the impact of the fastMRI dataset on the field of medical imaging is unquestioned, the dataset currently lacks clinical expert pathology annotations, critical to addressing clinically relevant reconstruction frameworks and exploring important questions regarding rendering of specific pathology using such novel approaches. This work introduces fastMRI+, which consists of 16154 subspecialist expert bounding box annotations and 13 study-level labels for 22 different pathology categories on the fastMRI knee dataset, and 7570 subspecialist expert bounding box annotations and 643 study-level labels for 30 different pathology categories for the fastMRI brain dataset. The fastMRI+ dataset is open access and aims to support further research and advancement of medical imaging in MRI reconstruction and beyond.

**Background & Summary**

Magnetic resonance imaging (MRI) is a widely utilized medical imaging modality critically important for a broad range of clinical diagnostic tasks including stroke, cancer, surgical planning, acute injuries, and more. Machine learning (ML) techniques have demonstrated opportunities to improve the MRI diagnostic workflow particularly in the image reconstruction task by saving time, reducing contrast, and leading in cases to FDA-cleared solutions[1-4]. Among the myriad applications of machine learning in medical imaging being explored, deep learning-based MRI reconstruction is one of the applications having the most impact on current clinical practice.

ML-based MRI reconstruction approaches often require data from "raw" fully sampled k-space datasets in order to generate ground truth images. However, large datasets of raw MRI measurements are generally not widely available. To address this need and facilitate cross-disciplinary research in accelerated MRI reconstruction using artificial intelligence, the fastMRI initiative was developed. fastMRI is a collaborative project between Facebook AI Research (FAIR), New York University (NYU) Grossman School of Medicine, and NYU Langone Health which includes the wide release of raw MRI data and image datasets[5]. While the fastMRI data has enabled exploration of ML-driven accelerated MRI reconstruction[6,7], there is a lack of clinical pathology information to accompany the imaging data which has limited the reconstruction assessment approaches to quantitative metrics such as structural similarity index measure (SSIM), leaving important questions regarding how various pathologies are represented in ML-based reconstruction unanswered[8].

In this paper, we present wide availability of a complementary dataset of annotations, fastMRI+, consisting of human subspecialist expert clinical bounding box labelled pathology annotations for knee and brain MRI scans from the fastMRI multi-coil dataset: specifically encompassing 16154 bounding box annotations and 13 study-level labels for 22 different pathology categories on knee MRIs, as well as 7570 bounding box annotations and 643 study-level labels for 30 different pathology categories for brain MRIs. This new dataset is open and accessible to all for educational and research purposes with the intent to catalyse new avenues of clinically relevant, ML-based reconstruction approaches and evaluation.

## Methods
### MRI image dataset

The fastMRI dataset is an open-source dataset, which contains raw and DICOM data from MRI acquisitions of knees and brains, described in detail elsewhere[5]. The images used in this study were directly obtained from the fastMRI dataset, reconstructed from fully sampled, multi-coil k-space data (both knee and brain). Image reconstruction was performed by inverse Fast Fourier Transform of each individual coil and coil combination with root sum square (RSS) for the purposes of creating fastMRI+. The reconstructed images were subsequently converted to DICOM format for human expert reader (radiologist) annotation.

### Annotations

Annotation was performed using a commercial browser-based annotation platform (MD.ai, New York, NY) which allowed adjustment of brightness, contrast, and magnification of the images. Readers used personal computers to view and annotate the images using the annotation platform.

A subspecialist board certified musculoskeletal radiologist with 6 years in practice experience performed annotation for the knee dataset and a subspecialist board certified neuroradiologist with 2 years in practice experience performed annotation for the brain dataset. Annotation was performed with bounding box annotation to include the relevant label for a given pathology on a slice-by-slice level. When more than one pathology was identified in a single image slice, multiple bounding boxes were used.

All 1172 fastMRI knee MRI raw dataset studies were reconstructed and clinically annotated for fastMRI+. Each knee examination consisted of a single series (either proton density (PD) or T2-weighted) of coronal images where bounding box labels were placed on each slice where representative pathology was identified[9,10]. Effort was made to try to include all the pathology within the bounding box while limiting the normal surrounding anatomy. If the examination contained significant clinically limiting artifacts, then the annotation for "Artifact" was added as a study-level label. In these instances, an interpolation tool was used in which the first and last slice were each labelled and the user interface interpolated the labels on intervening slices. If no relevant pathology was identified on an examination, no labels were provided.

A sub selection of 1001 out of 5847 fastMRI brain MRI raw dataset studies were selected for annotation. Each brain examination included a single axial series (either T2-weighted FLAIR, T1-weighted without contrast, or T1-weighted with contrast) where bounding box labels were placed on each image in which representative pathology or normal anatomical variant was identified[11,12]. As in knee examinations, effort was made to try to include all the pathology within the bounding box while limiting the normal surrounding anatomy. In some cases, the pathology or normal anatomic variant displayed within a given examination was so extensive or diffuse that a study-level label was used to characterize the relevant images or the entire exam inclusive of the finding (i.e., diffuse white matter disease). The study-level label, in these instances, replaced the use of a bounding box. If no relevant pathology was identified on a given examination, no labels were provided.

Note there are several limitations to this dataset that bear acknowledgement. First, while the annotators are subspecialist radiologists in practice at leading academic medical centers, the lack of multiple annotators to determine inter-rater reliability metrics or ensure consensus agreement is a limitation and should be considered in the use of these labels. Further work may include multiple annotations by multiple readers to further refine the clinical labels applied in fastMRI+. Additionally, the fastMRI knee MRI raw dataset contained only coronally acquired series while the brain MRI dataset contained only axially acquired series, each in a variety of pulse sequences and coils. While sufficient for annotation, it is important to note that true diagnostic interpretation in MRI for the included pathologies typically demands multi-sequence and multi-planar images for clinically accurate interpretation. Thus, the annotations provided by fastMRI+ may be incomplete. In the future, raw MRI datasets containing fully sampled multi-planar and multi-sequence data would enable optimal clinical annotation.

**Statistical Analysis**

Label distribution analysis was conducted for both knee and brain datasets showing detailed label descriptions at the same time. Table 1 shows annotation count and subject count for corresponding image-level knee labels. Note 'Artifact' is a study-level label for the entire study rather than a label of individual images. Table 2 shows annotation count and subject count for corresponding image-level brain labels. Table 3 shows subject count for corresponding subject-level brain labels.

Table 1. Knee label summary.

| Label | Annotation Count | Subject Count |
|---|---|---|
| Meniscus Tear | 5658 | 663 |
| Displaced Meniscal Tissue | 232 | 56 |
| Bone-Subchondral Edema | 986 | 196 |
| Bone Lesion | 183 | 29 |
| Bone-Fracture/Contusion/Dislocation | 1060 | 119 |
| ACL High Grade Sprain | 678 | 101 |
| ACL Low-Mod Grade Sprain | 765 | 153 |
| MCL High Grade Sprain | 11 | 4 |
| MCL Low-Mod Grade Sprain | 285 | 121 |
| PCL High Grade Sprain | 18 | 3 |
| PCL Low-Mod Grade Sprain | 142 | 40 |
| LCL Complex High Grade Sprain | 14 | 3 |
| LCL Complex Low-Mod Grade Sprain | 130 | 48 |
| Cartilage Full Thickness Loss/Defect | 615 | 122 |
| Cartilage Partial Thickness Loss/Defect | 2985 | 588 |
| Joint Effusion | 1311 | 142 |
| Joint Bodies | 38 | 11 |
| Periarticular Cysts | 864 | 161 |
| Muscle Strain | 65 | 11 |
| Soft Tissue Lesion | 90 | 10 |
| Patellar Retinaculum High Grade Sprain | 24 | 4 |
| Artifact | / | 13 |

*Artifact is study-level label

Table 2. Brain image-level label summary

| Image Level Label | Annotation Count | Subject Count |
|---|---|---|
| Absent Septum Pellucidum | 3 | 1 |
| Craniectomy | 32 | 4 |
| Craniotomy | 1025 | 99 |
| Craniotomy with Cranioplasty | 43 | 3 |
| Dural Thickening | 351 | 30 |
| Edema | 369 | 44 |
| Encephalomalacia | 161 | 18 |
| Enlarged Ventricles | 300 | 38 |
| Extra-Axial Mass | 104 | 11 |
| Intraventricular Substance | 8 | 1 |
| Likely Cysts | 17 | 5 |
| Lacunar Infarct | 113 | 32 |
| Mass | 380 | 46 |
| Nonspecific Lesion | 757 | 124 |
| Nonspecific White Matter Lesion | 1826 | 173 |
| Normal Variant | 73 | 21 |
| Paranasal Sinus Opacification | 40 | 8 |
| Pineal Cyst | 2 | 1 |
| Possible Artifact | 505 | 52 |
| Posttreatment Change | 1262 | 99 |
| Resection Cavity | 199 | 27 |

*Likely Cysts is applied to small lesions (approximately 1 cm or less in diameter) which are difficult to distinguish from parenchymal, simple parenchymal neuronal cyst, and prominent perivascular space.

Table 3. Brain study-level label summary

| Study Level Label | Subject Count |
|---|---|
| Global Ischemia | 1 |
| Small Vessel Chronic White Matter Ischemic Change | 221 |
| Motion Artifact | 33 |
| Possible Demyelinating Disease | 2 |
| Colpocephaly | 2 |
| White Matter Disease | 2 |
| Innumerable Bilateral Focal Brain Lesions | 2 |
| Extra-Axial Collection | 9 |
| Normal for Age | 371 |

**Data Records**

We created separate annotation files for the 1172 validation knee datasets and 1001 brain datasets, all based on the fastMRI source data[5]. The annotation files (knee.csv and brain.csv) can be accessed from fastMRI-plus GitHub repository (https://github.com/microsoft/fastmri-plus) in CSV formats. Four CSV files are included in the 'Annotations' folder. File names of all radiologist-interpreted dataset are stored in knee_file_list.csv and brain_file_list.csv, respectively. Annotations are contained in knee.csv and brain.csv. In each annotation CSV file, the file names (i.e., column 'Filename') are aligned with the naming in the fastMRI dataset. For each annotation, the file name, slice number, bounding box information, and disease label are provided. The bounding box information includes four parameters, x, y, width (pixel), and height (pixel), representing the x and y coordinates of the upper-left corner, the width and height of the bounding box. Unit of the bounding box parameters is 'pixel'. Study-level labels are marked as 'Yes' in column 'Study Level' for slice 0 of the corresponding subjects with no specified bounding box information.

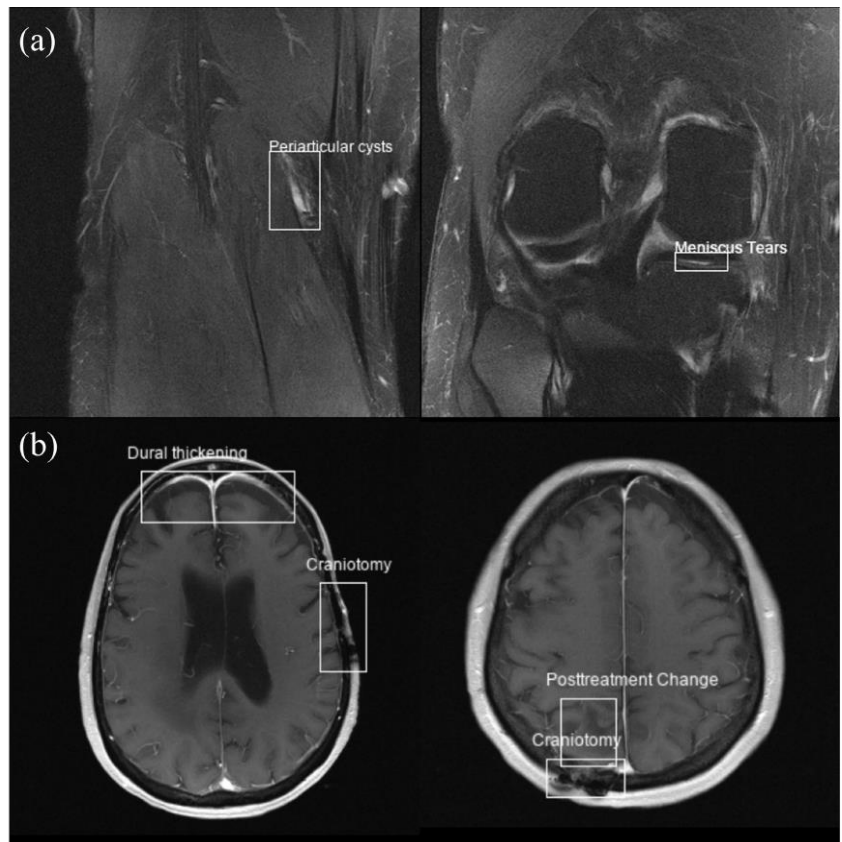

Figure 1. Example annotations (labels and bounding boxes) from the fastMRI+ dataset shown superimposed on both knee (a) and brain (b) reconstructed images from the fastMRI dataset open-source repository.

**Usage Notes**

The bounding box information can be used to plot overlaid bounding boxes on images, as shown in Fig.1. The clinical labels, together with the bounding box coordinates, can also be converted to other formats (e.g., YOLO format[13]) in order to configure a classification or object detection problem. The open-source repository also contains an example Jupyter Notebook ('ExampleScripts/example.ipynb') of how to read the annotations and plot images with bounding boxes in Python.

**Code Availability**

**Acknowledgements**

The authors want to thank Desney Tan at Microsoft Research and Michael Recht at New York University for project support. Sincerest thanks to George Shih and Quan Zhou at MD.ai for providing annotation infrastructure.

**Author contributions**

R.Z., B.Y., and Y.Z. contributed equally to data processing, data analysis, and manuscript preparation. R.S. and A.D. contributed to data annotation work. F.K., Z.H., Y.L. authorized and facilitated access and usage of raw fastMRI dataset and contributed to manuscript editing. M.S.H. and M.P.L. coordinated and led all details of this project, manuscript composition, and editing.

**Competing interests**

The authors declare no competing interests related to this work.

**Figure Legends**

Figure 1. Example annotations (labels and bounding boxes) from the fastMRI+ dataset shown superimposed on both knee (a) and brain (b) reconstructed images from the fastMRI dataset open-source repository.